# Tropospheric Ozone Formation Potential and Related Design Considerations for Radiative Coolers

by

Jyothis Anand[1], Piero Di Carlo[2], Eleonora Aruffo[3]*, Paul E. Ohno[4]*, Jyotirmoy Mandal[5]*

(1) Oak Ridge National Laboratory, Oak Ridge, TN 37830, USA

(2) Department of Advanced Technologies in Medicine & Dentistry, University "G. d'Annunzio" of Chieti-Pescara, Chieti 66100, Italy

(3) Department of Science, University "G. d'Annunzio" of Chieti-Pescara, Chieti 66100, Italy

(4) Department of Chemistry and Biochemistry, Auburn University, Auburn, AL 36849, USA

(5) Department of Civil and Environmental Engineering, Princeton University, Princeton, NJ 08544, USA

*To Whom Correspondence Should be Addressed

Email: eleonora.aruffo@unich.it, pohno@auburn.edu, jm3136@princeton.edu

**Abstract**

In recent years, solar reflective radiative coolers have been proposed as a passive cooling design: both for reducing cooling loads and lowering temperatures in buildings[1], and at urban scales as a heat mitigation measure.[2, 3] However, the potential impacts of widespread deployment of radiative coolers on the chemical composition of the atmosphere remain largely unexplored. A defining feature of recently-designed highly solar reflective radiative coolers is their high ultraviolet (UV) reflectance, which is required for sub-ambient cooling under strong sunlight. Yet, wide adoption of such UV-reflective radiative coolers could substantially increase the UV actinic flux in the atmosphere above. This, in turn, may affect tropospheric ozone concentrations, particularly in urban atmospheres with high $NO_x$ concentrations. Here, as a case study, we use a 0-dimensional photochemical box model, constrained by field measurements of meteorological conditions and chemical concentrations in the urban environment of Houston, Texas , to explore the potential impact of the widespread use of UV-reflective radiative coolers on ozone concentrations. The average surface albedo is modified to represent citywide rooftop deployments of different passive cooling designs. The effects of changing surface albedo on UV and visible flux is then calculated using a radiative transfer model and the photochemical box model is used to determine the impacts of these flux changes on ozone concentrations. Our calculations show that complete deployment of radiative coolers may increase tropospheric ozone levels by as much as 30% during specific meteorological conditions in Houston. Informed by the wavelength-dependent modelling results, we propose specific designs – namely pigmented radiative coolers with different UV reflectances, and UV-absorptive visible-reemitting fluorescent radiative coolers – that could minimize negative ozone formation while retaining appreciable cooling performance. Our results motivate further study on the effects of widespread deployment of radiative cooling designs like superwhite roof paints on air quality, and materials that simultaneously minimize adverse photochemical impact and maximize cooling performance.

## Introduction

Anthropogenic climate change necessitates the deployment of mitigation and adaptation measures to counter its widespread deleterious societal effects. Specifically exacerbating the impact of the associated average global temperature increase for city dwellers, the urban heat island effect is caused by the urban environment generally having a lower shortwave albedo and evotranspiration compared to natural environments as vegetation is replaced with heavy absorbers such as pavement and other roofing materials. One proposed way to counteract increased temperatures related to both climate change and urban heat islands is to replace traditional roofs and other low albedo materials with higher albedo counterparts. These replacements range from white paints[4] to advanced engineered materials with carefully designed reflectance and emittance characteristics. Some of these materials, termed radiative coolers,[5-8] can achieve sub-ambient temperatures even under direct sunlight. This effect is achieved through efficient skyward broadband thermal infrared ($\lambda{\sim}2.5 - 30\ \mu m$) emission, especially through the atmosphere's longwave infrared ($\lambda{\sim}8 - 13\ \mu m$) transmission window, and high reflection of the incident solar wavelengths ($\lambda{\sim}0.3 - 2.5\ \mu m$). Depending on the specific design, this high reflectance extends to a varying degree down into the UV region ($\lambda{\sim}0.3 - 0.4\ \mu m$) and in some cases can result in near-unity solar reflectances.

Because many of these radiative cooler designs have only been proposed and experimentally realized within the past decade, the potential impacts of their widespread deployment on the air quality of the urban environment remain incompletely understood. An increased UV flux due to an increased UV reflectance by the earth's surface can impact key photochemical reactions that play a large role in governing the concentration of ozone and other trace species of note in the troposphere. Ozone photolysis is one such reaction:

$$O_3 + h\nu \rightarrow O_2 + O(^1D) \quad (1)$$

Since radiation of wavelengths $\lesssim$ 290 nm does not reach the troposphere and the ozone photolytic cross section and $O(^1D)$ quantum yield both fall off rapidly above 300 nm, 290-320 nm is the relevant wavelength range for reaction 1.[9, 10] A collision of the excited oxygen atom produced in reaction 1 with a water molecule leads to the production of two OH radicals. OH radicals are the predominant oxidant in the troposphere during the day, and their reactions with volatile organic compounds (VOCs) kick off a complex network of oxidation reactions, the exact course of which depend on the relative concentrations of a variety of trace constituents in the atmosphere.

A second key photochemical reaction in the troposphere is the interconversion of $NO_x$ through the photolysis of $NO_2$:

$$NO_2 + h\nu \rightarrow NO + O \tag{2}$$

The relevant wavelength range in the troposphere for Reaction 2 begins similarly at the shortest wavelengths that reach the troposphere and continues to approximately 410 nm where the quantum yield for photolysis begins to approach zero.[9, 10] Reaction 2 is the dominant production pathway for ozone in the troposphere. This reaction results in ozone formation upon the reaction of the produced oxygen atom with molecular oxygen.

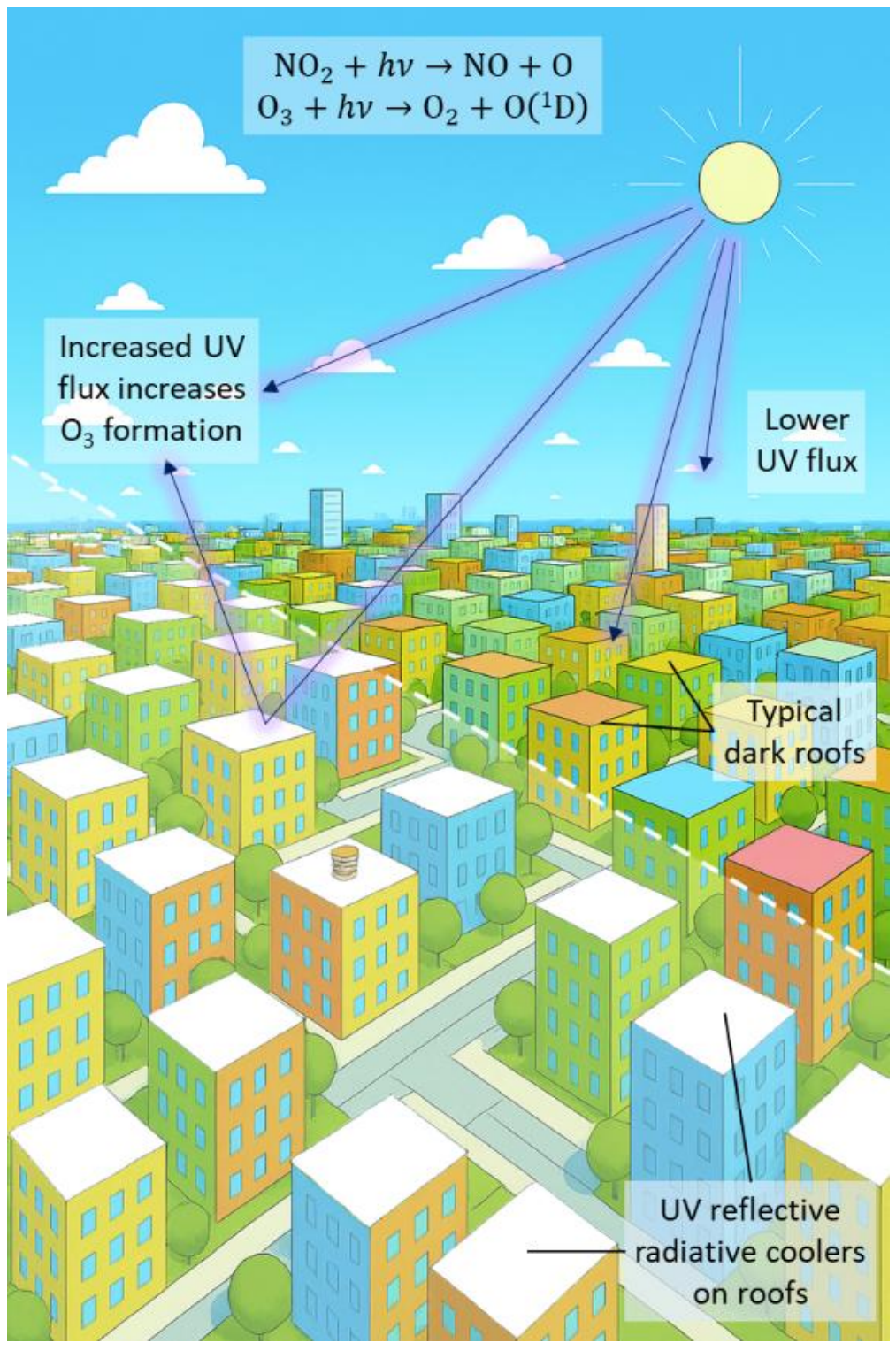

Exposure to elevated ozone concentrations has been shown to exacerbate lung and respiratory and lung diseases such as asthma, emphysema, and chronic bronchitis, and extensive epidemiological evidence indicates that many thousands of premature deaths each year are attributable to ozone exposure.[11, 12] While it is intuitive that changing UV flux can affect tropospheric ozone concentrations by altering rates of photochemical reactions including 1 and 2 as illustrated schematically in *Figure 1*, the complexity of the interconnected network of reactions and reactants connecting these reactions and others to ozone concentrations necessitates the use of a model chemical mechanism to quantify these effects.

***Figure 1****: Schematic illustration of the impact of radiative cooler deployment on tropospheric chemistry*

Epstein et al.[13] investigated the air quality impacts of high-albedo roofing materials in Southern California using the Community Multiscale Air Quality model and found multiple competing pathways affecting ozone concentrations: ozone decreases from temperature reduction, ozone increases from reduced mixing due to decreases in boundary layer height, and ozone increases from increased UV flux, with the net effect ultimately determined by the detailed meteorological conditions and materials deployed. The first two pathways are directly related to the net temperature impacts of high albedo materials and have been the topic of previous study.[14-16] However, highly UV-reflective radiative coolers have not been studied in these contexts. Here, we focus specifically on the final pathway as it has the possibility to be an additional design parameter around which competing radiative cooler designs can be balanced and optimized in addition to their cooling performance.

We investigate the impacts on ozone concentrations of the widespread deployment of radiative cooler materials with varying wavelength ($\lambda$) -dependent reflectance using the zero-dimensional (0-D) photochemical box model F0AM and the Master Chemical Mechanism (MCM), constrained by field observations in Houston, TX. We model the change in albedo that would result from the deployment of each type of material: titanium dioxide ($TiO_2$)-based white paints that absorb at $\lambda \lesssim 410$ nm, silvered materials that begin to absorb at $\lambda \lesssim 330$ nm, and superwhite materials that reflect all the way down to the cutoff at $\lambda \sim 290$ nm beyond which solar radiation is essentially all absorbed prior to reaching the troposphere. Using a radiative transfer model, we calculate the change in actinic flux that would result from these changes in albedo and use F0AM to calculate resulting ozone concentrations. While this simplified approach does not consider complex coupled interactions between meteorology and chemistry, nor the effects of transport of reactive species, it nonetheless allows the isolation and exploration of the effect that changing UV albedo in different bands has on this complex chemistry in order to gain conceptual understanding of the wavelength-dependent formation and loss mechanisms of ozone under these conditions.

Exploration of the spectral dependence of $NO_x$ photolysis can also shed light on radiative cooler design. We first explore the tradeoff between cooling performance and adverse photochemical impact through ozone formation as a function of the UV reflectance of radiative coolers. Subsequently, we explore how part of the absorbed UV radiation can be fluoresced back skywards in the visible wavelengths, thus increasing the effective reflectance. Subsequently, we theoretically explore how real radiative coolers containing pigments with different band gaps in the UV can yield various points in the tradeoff between reflectance and photochemical impact, and how a UV-absorptive, blue-fluorescent bilayer radiative cooler designs comprisable from reported materials could largely prevent an increase in ozone formation while maintaining a high solar reflectance. Collectively, these explorations highlight new considerations for radiative cooling research, and a new criteria for radiative cooler design.

## Modelling the Impact of Radiative Coolers on Atmospheric Ozone

Tropospheric photochemistry was simulated using the Framework for 0-D Atmospheric Modeling (F0AM), a 0-D photochemical box model written in MATLAB.[17] The simulations were constrained using measurements taken during an urban field campaign in Houston, TX as a case study for an urban environment with relatively high $NO_x$ and VOC concentrations. Field observations of ozone, CO, $SO_2$ and VOC concentrations, as well as other meteorological variables were made at the La Porte Municipal Airport (95°03’51.1’’W, 29°40’09.3’’N, altitude: 8 m asl), Texas, from August 15th to September 15th 2000, during the Texas Air Quality Study 2000. A complete description of the field campaign, including the instrumentation and acquisition methods, can be found in the literature.[18] The airport is primarily used by sport airplanes during weekends and local flight schools for their training throughout the week. The site is surrounded by a number of power plants and petrochemical industries that characterize the environment. Intense traffic on the highways in the vicinity of the city could also impact the site. The sea breeze with its sea-land circulation breaks up stagnant air masses.

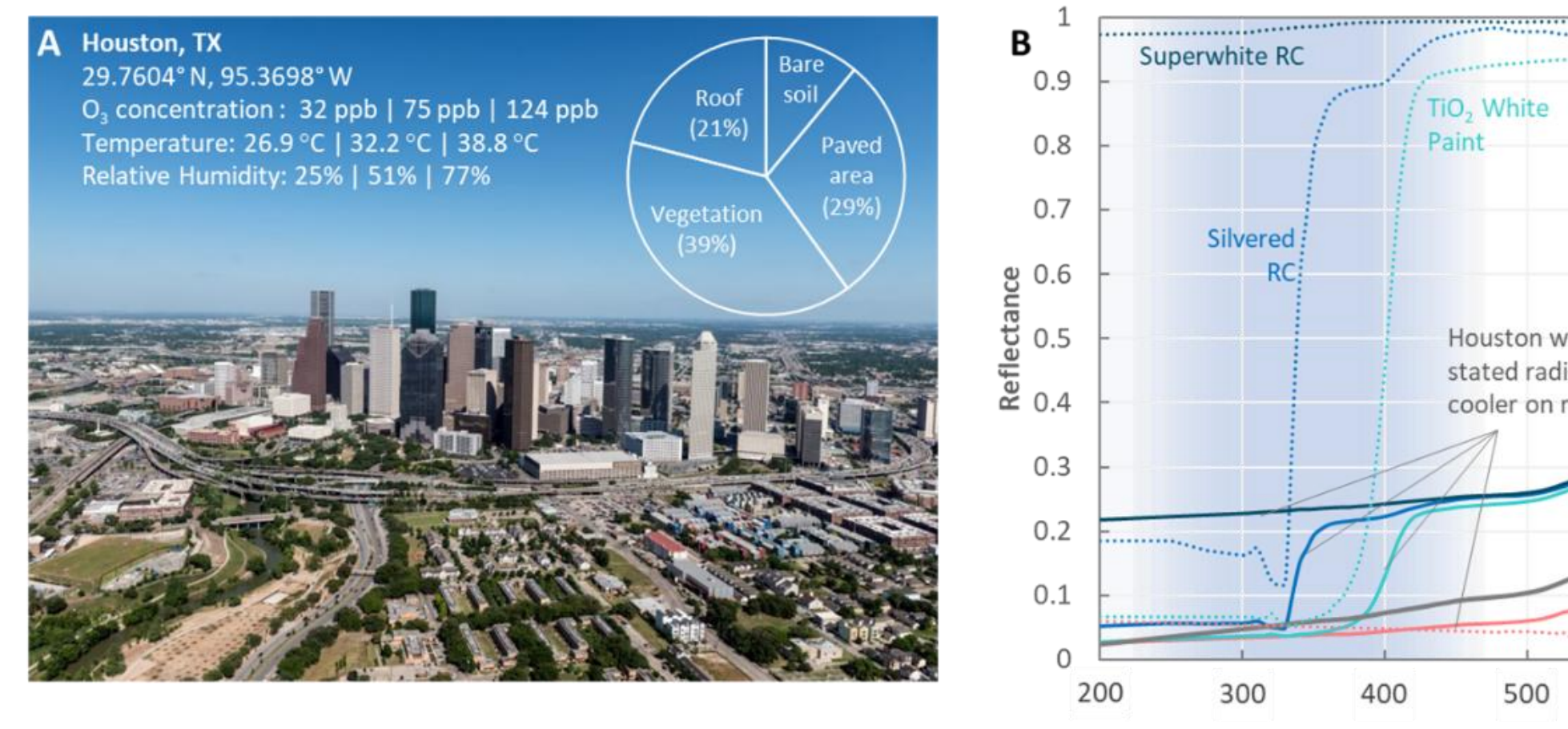

***Figure 2****: (A) Breakdown of Houston land area used in the model scenarios. Meteorological data is presented for the field campaign duration, and in percentile values* ($P_{10}$| $P_{50}$| $P_{90}$). *(B) Spectral reflectance of each roof material we considered (dotted lines) and the calculated average albedo for each scenario (solid lines).*

Observed chemical concentrations and meteorological measurements of pressure, temperature, and relative humidity were used as F0AM model inputs. The model run consisted of individual steps of duration 7200 s with a dilution term of 1/86400 $s^{-1}$. The concentrations of the experimentally measured compounds and meteorological parameters were held constant throughout the duration of each step. The ozone concentration was analyzed as the model output for all the different scenarios simulated. Gaps in experimentally measured data were not imputed and only the data points in which all the species serving as model inputs were available were included in the model run. The experimentally measured species used to constrain the box model were: OH, $HO_2$, $NO$, $NO_2$, $SO_2$, $C_5H_8$, α-pinene, isopentane, n-pentane, n-hexane, octane, benzene, toluene, ethylbenzene, m-xylene, o-xylene, acrolein, acetaldehyde, acetone, methacrolein, methyl vinyl ketone, and methyl ethyl ketone.

Photochemistry was handled using the “bottom-up” method of *j*-value (photolysis rate constant) calculation within F0AM. This approach combines a user-specified actinic flux spectrum with literature-derived cross sections and quantum yields. Spectral actinic fluxes and associated *j*-values were calculated using the Tropospheric Ultraviolet and Visible (TUV) Radiation Model from NCAR, Version 5.4. Calculated actinic fluxes were corrected for variability in clouds or other atmospheric variables by multiplying them by the measured/calculated *j*-value ratio of $NO_2$ photolysis (reaction 2), for which the *j*-value was measured experimentally during the campaign. The TUV subroutine ‘setalb’ was modified so that a spectrally-resolved albedo could be used in the model. The latitude for Houston was set at 29.760° and longitude at -95.730°. The 4 stream pseudo-spherical discrete ordinate radiation transfer model was used. Spectral actinic flux at 10 m altitude was calculated from 206 to 850 nm. For the calculation, it was assumed that there was no cloud cover and that the aerosol vertical optical depth was 0.235 with an aerosol single scattering albedo of 0.99; variability in these parameters was assumed to be corrected for by the previously mentioned experimental *j*-value correction.

The average albedo for each scenario was calculated by accounting for the area fractions of different materials in the Houston area, taken from the literature as: barren land (11%), vegetation (39%), paved areas including roads, pavements and parking lots (29%), and roofs (21%).[19] The spectral reflectance of barren land, vegetation, and paved areas (assumed to be concrete) were obtained from the literature.[20] The baseline scenario was assumed to have all roofs having the optical properties of concrete. The superwhite radiative cooling paint was assumed to be a previously reported porous P(VdF-HFP) coating,[7] the silvered radiative cooler was assumed to be a silvered silica polymer composite film,[5] and the traditional white paint was assumed to be a commercial TiO2-based paint from Sherwin Williams.[7] Reflectance of the black paint was obtained by measuring a carbon-black based variant. All the measurements were taken at near normal incidence. It should be noted that the data for vegetation, paved areas and soil have large uncertainties owing to intrinsic variations. However, this occurs in the longer wavelengths where they have higher reflectances. In the shorter violet to UV wavelengths that are of consequence to ozone formation, their reflectances are low, and the associated uncertainties small.

Figure 2 shows the different albedo scenarios used in this study. The baseline case was calculated assuming average reflectances for each category of land in Houston (barren land, vegetation, pavement, and roofs), the percentages of which are shown in Figure 2a. As a limiting case, in each of the perturbed scenarios it was assumed that the entirety of the roofs in Houston were replaced with the material specified. The 'black paint' scenario was used to investigate the hypothetical effect of decreasing albedo. Three different scenarios with an increase in albedo over the baseline case were investigated. The '$TiO_2$' scenario represents the case where standard $TiO_2$-based white paints are applied on all roofs . These paints are characterized by a sharp decrease in reflectance at approximately $\lambda \sim 400$ nm. The 'silvered' case reflects the deployment of silvered materials that extend their high reflectance into the UV but still drop off sharply at approximately $\lambda \sim 400$ nm. The 'superwhite' case reflects the use of superwhite materials that maintain a high reflectance throughout the photochemically active UV region.

### Effect of albedo changes on ozone concentrations.

Figure 3a-e show the model performance for each case with the measured ozone concentration on the x-axis and the modeled concentration on the y-axis for each time point. Figure 3a shows the baseline case. The slope of the line of best fit of 0.97 shows that model is not substantially systematically over- or under-predicting the ozone concentration, and that our choice of reflectances for the different land surface components may be reasonable. The $R^2$ value of 0.80 shows that the model is adequately representing the overall chemistry. Because the model necessarily includes only a subset of the chemistry that occurs in the troposphere and the experimental data only includes a small fraction of species, an inability to perfectly recapitulate the measured ozone concentrations with the model is expected. Notably, correcting the modeled *j*-values for the cloud cover by normalization with the observed *j*-values was found to be critical for model performance, increasing the $R^2$ value from 0.65 without the correction to 0.80 with this correction.

Figure 3b shows the model results using the albedo resulting from the replacement of roofs with black paint. The slope of the line of best fit of 0.93 shows a modest decrease in predicted ozone concentrations compared to the baseline case (slope = 0.97). Such a decrease is expected based on the decrease in albedo causing an associated decrease in UV flux in the troposphere. Figures 3c-e show the model results of the three cases representing the deployment of differing radiative cooler materials. The '$TiO_2$' white paint case is shown in Figure 3c, showing a slope of 0.98, similar to that of the baseline case. This similarity can be rationalized by the fact that $TiO_2$-based white paint is a modest radiative cooler due to its high reflectance in the visible region but critically this high reflectance does not extend to the UV wavelengths associated with photochemical reactions 1 and 2. (Figure 2b).

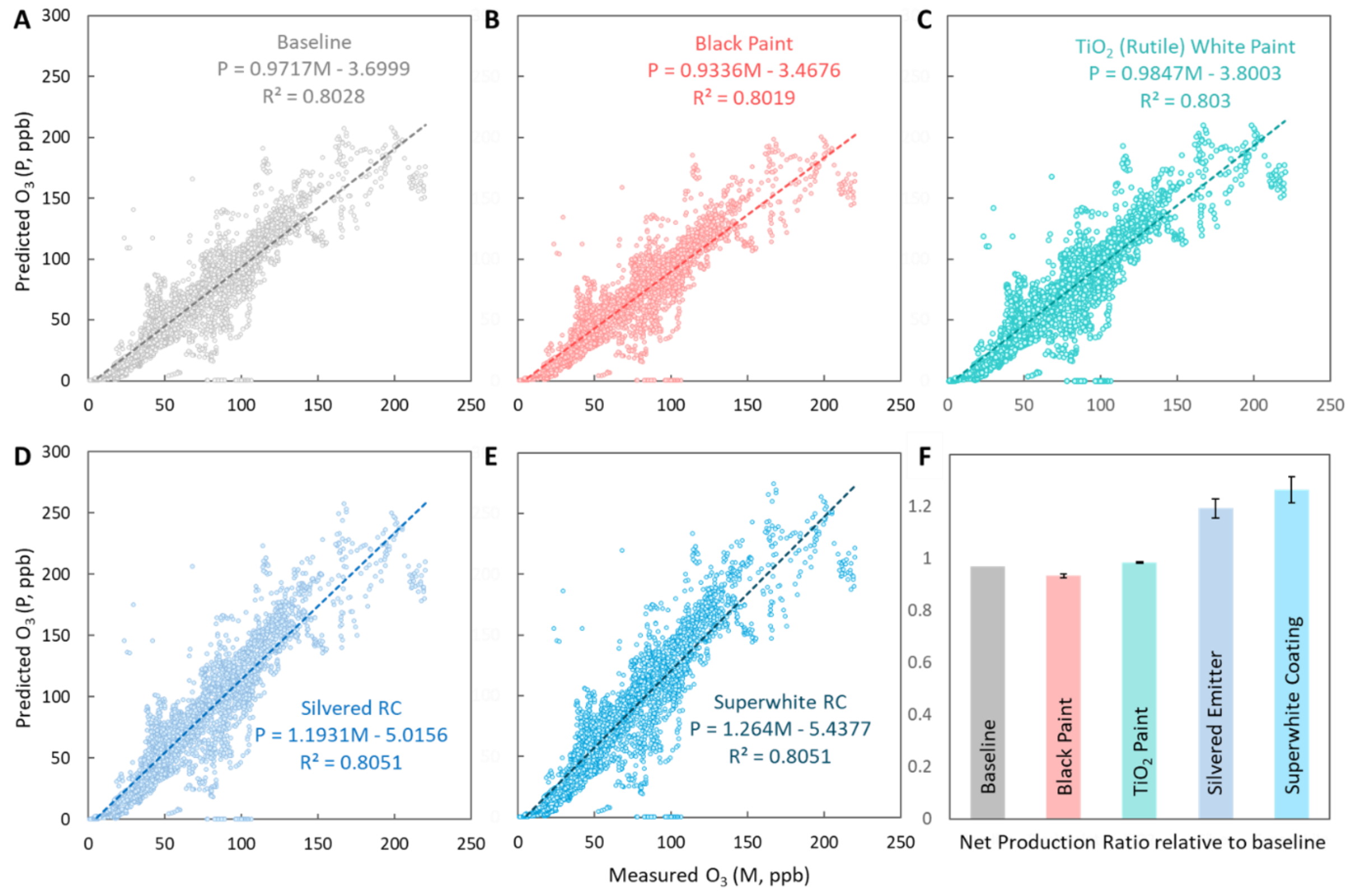

***Figure 3**: Modeled ozone concentrations upon albedo perturbation vs measured ozone concentrations for (a) baseline, (b) black paint, (c) TiO2 white paint, (d) silvered emitter, and (e) superwhite scenarios. (f) shows the slope of the line of best fit of the predicted vs measured plots for each scenario.*

In contrast, Figures 3d and 3e show the simulated results representing the deployment of radiative coolers that are reflective in this UV range and thus do increase the flux in this critical wavelength region. Figure 3d shows the case of silvered radiative coolers, resulting in a slope of 1.19, a 23% increase compared to the baseline case. Figure 3e shows the case of the Superwhite radiative coolers, resulting in a slope of 1.26, a 30% increase compared to the baseline case. The model thus predicts that the increased UV-flux resulting from the widespread deployment of these coolers would lead to an associated increase in ozone in the troposphere. Notably, despite a lower albedo throughout the UV and an absorbance cutoff around 340 nm, the silvered case produces nearly 75% of the ozone increase as that of the full superwhite case. Despite the silvered cutoff at 340 nm, these materials still exhibit enhanced reflectance for the majority of the relevant wavelength range for Reaction 2, the dominant formation mechanism of ozone in the troposphere. Figure 3f shows a summary of the results shown in Figures 3a-e. We note that the model does not take into account complex coupled meteorology and chemistry, though a separate WRF simulation (SI, Section S1) suggests that the change in temperature produced by the radiative cooler scenarios explored here would only minorly offset the increase ozone production found in here.

The main production reaction for ozone formation is $O + O_2 \rightarrow O_3$. In the troposphere, the O atoms are mainly generated from photolysis of $O_3$ (a null cycle) or photolysis of $NO_2$. Once formed, $O_3$ reacts with NO to regenerate $NO_2$. In a simplified analysis of the numerous coupled reactions that are treated in the F0AM model, a basic analysis using only the previous reaction cycle can be performed. The steady-state approximation can be-invoked regarding the oxygen atom due to its high reactivity, which leads to the following steady-state ozone concentration, known as the photostationary state relation[9]:

$$[\mathrm{O_3}] = \frac{j_{\mathrm{NO_2}}[\mathrm{NO_2}]}{k_{\mathrm{O_3+NO}}[\mathrm{NO}]}$$

where $j$ represents a photolysis rate constant and $k$ represents and non-photolysis chemical rate constant. This simplified relationship, which assumes no other source or loss of the species present besides the reactions previously mentioned, demonstrates a simplified mechanism through which the increased UV flux (and thus $j_{\mathrm{NO_2}}$) can lead to an increase in $O_3$ concentrations. In actuality, the model used here also includes organonitrate chemistry, which involves numerous coupled reactions between VOCs, $NO_x$, $HO_x$, and $O_3$.

## Implications for Radiative Cooler Design

The model results shown in Figure 3 can be used to guide the design of radiative coolers to mitigate increase in ozone concentrations while retaining cooler potential. The key wavelengths for Reactions 1 and 2 can be determined by multiplying the actinic flux by the reaction cross section and quantum yield for each wavelength. The actinic flux, individual cross sections, and quantum yields can be found in the SI, Figures S2-S4. Results for the baseline and superwhite cases are shown in Figure 4 for Reaction 1 (Figure 4a) and Reaction 2 (Figure 4b). Figure 4a shows that Reaction 1 is triggered by a narrow wavelength range peaked sharply at $\lambda \sim 308$ nm. In contrast, Reaction 2 is triggered by a broad wavelength range of $\lambda \sim 300$–$410$ nm, albeit with photons at the longer-wavelength end contributing more to the reaction events. Crucially, this increase is caused by not only increasing flux from $\lambda \sim 350$–410 nm, but also by increasing $NO_2$ cross section at the longer wavelengths in this region. Thus, a radiative cooler could in principle be designed to balance increased ozone formation with cooling performance by not exhibiting a high reflectance in this band.

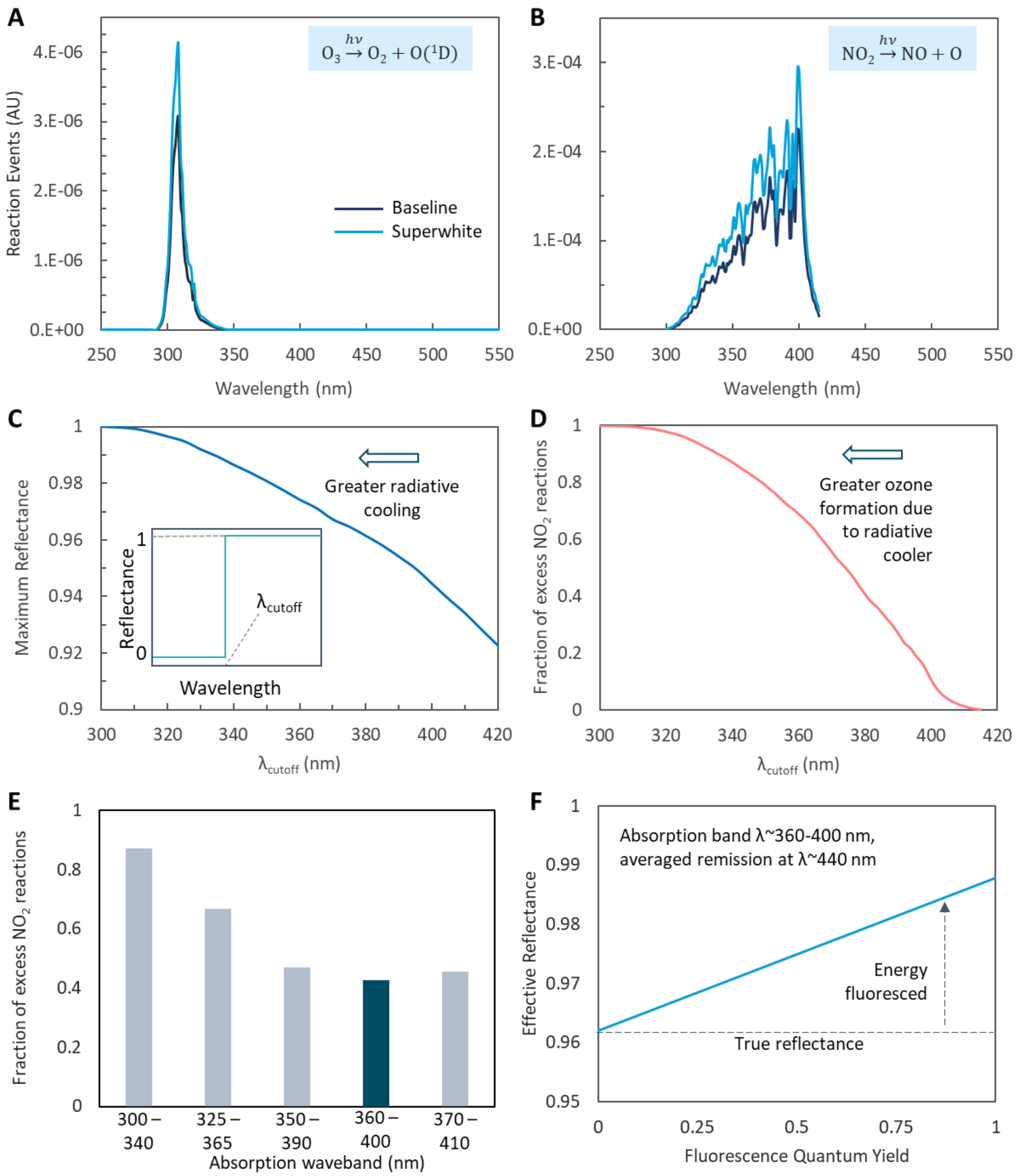

***Figure 4***. *Wavelength-dependence of photolysis reaction events for (**A**) $O_3$ photolysis producing $O(^1D)$ (**B**) $NO_2$ photolysis (**C**) The maximum reflectance of a radiative cooler as a function of shortwave reflectance cutoff wavelength (shown in the inset), and corresponding fraction of excess $NO_2$ photolysis reaction events. (**D**) Fraction of excess NO2 photolysis events for hypothetical ideal reflectors with cutoffs, relative to a perfectly reflective radiative cooler, as a function of the cutoff. (**E**) Fraction of excess $NO_2$ photolysis events, relative to a perfectly reflective radiative cooler, for fluorescent materials with*

*different excitation bands and perfect reflectance elsewhere. Reemission is assumed to be at wavelengths > 410 nm. (**F**) For a fluorescent agent with excitation band of 360-400 nm and average reemission at 440 nm, the effective reflectance increases linearly with quantum yield.*

A consideration of the number of photolysis reactions events triggered by photons of each wavelength allows us to quantitatively investigate this phenomenon. To do so, we first considered a hypothetical ideal radiative cooler with a short wavelength reflectance of 0, a long wavelength reflectance of 1 up to λ~2500 nm, and a step change between these two reflectances at a specified cutoff wavelength in the UV-to-violet waveband (310-420 nm). Figure 4C shows the overall solar reflectance for these materials as a function of the cutoff wavelength. Figure 4D shows the fraction of excess NO2 photolysis events from reflected photons. As shown, as the reflectance cutoff moves from the edge of the solar spectrum to longer wavelengths, both total solar reflectance and NO2 photolysis events decrease. The steepest decrease in NO2 photolysis events occurs in the 350 - 400 nm range. While an 'ideal' reflectance may be contextual (e.g. based on cooling needs vs air quality concerns), Figure 4C-D demonstrates that it is possible that radiative coolers could be designed to balance cooling efficiency and air quality impacts by selecting a cutoff wavelength.

Beyond simply absorbing ultraviolet sunlight, which may not be ideal in scenarios where cooling potentials are low and solar heating is undesirable, a second approach may be to fluoresce back some of the absorbed energy skywards in longer wavelengths, thus achieving an effective solar reflectance $R_{solar,effective}$ higher than the true reflectance $R_{solar}$ (SI, Section S2). This could in principle be achieved with a bilayer structure, comprising a highly broadband solar reflective underlayer, and a topcoat consisting of one or multiple dyes with absorption in the critical UV region. As a hypothetical design, we consider a fluorescent bilayer with a perfect absorption and excitation waveband over part of the UV-to-violet wavelengths, and perfect reflectance in all other solar wavelengths. Figure 4E shows the fraction of excess $NO_2$ photolysis events relative to perfectly black roofs for a set of such hypothetical materials. As can be rationalized through the data shown in Figure 4B, a minimum in excess $NO_2$ photolysis events, ~57%, is predicted for an absorption band in the 360 – 400 nm wavelength range. Critically, since the absorptive layer consists of fluorescent dyes, some of this radiation can be reradiated. Such reradiation, expected to occur outside of the excitation band due to typical molecular Stokes shifts, would increase the effective reflectance of the material with an associated thermalization corresponding to $h(\nu_{ex}-\nu_{em})$, where $\nu_{ex}$ and $\nu_{em}$ are the frequencies of excitation and emission. Figure 4F shows calculated effective solar reflectance as a function of the material fluorescence quantum yield. A hypothetical bilayer material consisting of an ideal reflector combined with an ideal absorber in the range of 360 – 400 nm with a fluorescence quantum yield of 1, and an average emission at $\nu_{em}$= 440 nm is predicted to result in a ~ 57% reduction in excess NO2 photolysis events while reducing $R_{solar,effective}$ by only ~0.01. While real-world materials will necessarily not attain such performances, Figure 4E-F nonetheless outlines the basic parameters of how such a bilayer radiative cooler material might perform.

To motivate physical realizations of the possibilities suggested in Figs. 4C-F, we conclude by exploring relevant materials platforms. As a simple approach, we suggest pigments with bandgaps that lie in the UV wavelengths. Potential examples include rutile $TiO_2$,[7] anatase $TiO_2$, ZnO, ZnS, $SrTiO_3$,[21] and $BaTiO_3$, which have different bandgaps in the violet-UV region and absorb near and above-bandgap light. Fig. 5A shows the spectral reflectance of powdered rutile $TiO_2$, anatase $TiO_2$, ZnO and $SrTiO_3$, compared to the UV-reflective silvered polymer, $BaSO_4$, and porous polymer-based radiative coolers discussed earlier. Figs. 5C and D show the associated reduction in solar reflectance $R_{solar}$ and the fraction of excess $NO_2$ photolysis events associated with each material, respectively. As can be seen, a material such as SrTiO3, with a cutoff wavelength of ~365 nm, may reduce excess $NO_2$ photolysis events to nearly half that of UV-reflective radiative coolers, while reducing $R_{solar}$ by ~0.03. By contrast, rutile $TiO_2$, which is already used in cool roof coatings, reduces $NO_2$ photolysis events by ~80%, but reduces $R_{solar}$ by > 0.05. It should be noted that under strong sunlight, a difference of 0.01 in $R_{solar}$ corresponds to ~10 $Wm^{-2}$ heating by the sun. The stated materials can be incorporated into both polymer-composite,[21, 22] and as Figs. 5C-D indicate, can achieve different tradeoffs between cooling performance and ozone reduction.

As a potential fluorescent design, we theoretically explore a bilayer design comprising a commercial fluorescent polymer (EJ-299-27, polyvinyltoluene-coated plastic, from Eljen Technology) coating above a highly solar reflective radiative cooler.[7] Based on commercial literature, the topcoat is assumed to absorb nearly all ultraviolet light and reemit it across the violet-blue-green wavelengths (Fig 5B, inset) with a 90% fluorescence quantum yield. The reemitted light either escapes skywards, or is reflected by the radiative cooler below. Calculations based on the optical parameters indicate that the design may absorb nearly all incident UV light and reduces excess $NO_2$ photolysis and solar reflectance by 84% and ~5% respectively, much like $TiO_2$ (Fig. 5C-D). However, the fluorescence causes a significant fraction of the absorbed UV solar energy to be reemitted skywards as visible light, resulting in a higher effective solar reflectance at each absorbed wavelength (Fig. 5B, SI, Section 2), and consequently, a much smaller reduction of $R_{solar,effective}$ (< 0.02) than that for $R_{solar}$ (~0.05) (Fig. 5C). This is close to our calculated results in Fig. 4F.

Although our calculations are theoretical and based on potentially high-end fluorescent and radiative cooling materials, the potential performance, and the fact that such materials are commercially available, may motivate practical designs that optimize for both cooling and near-surface ozone reduction. For instance, the widespread usage of UV absorptive blue-fluorescent agents for brightening paper indicate that such designs may be achievable at scale.[23]

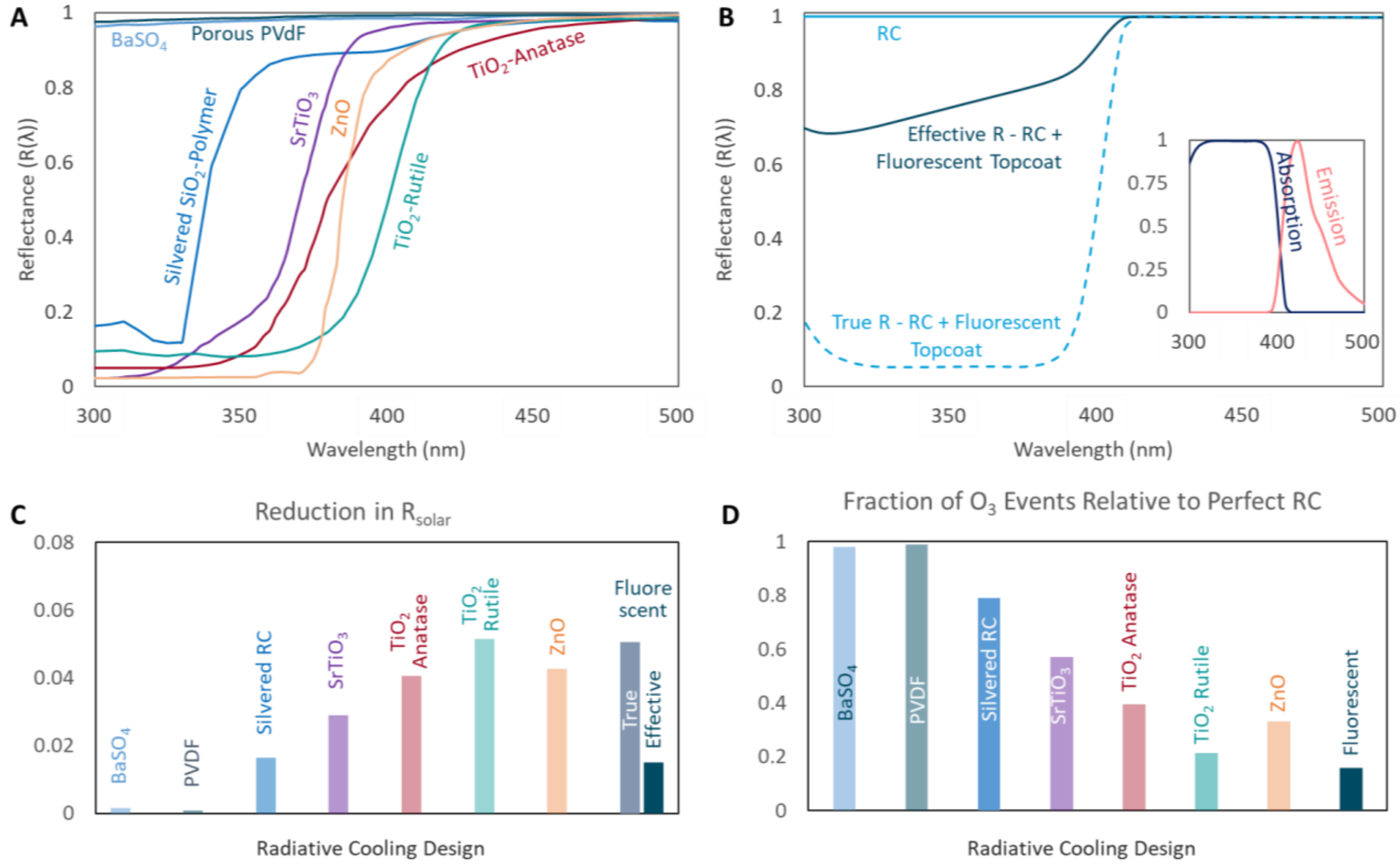

***Figure 5**. (**A**) Reflectances of different radiative cooling materials with characteristic levels of UV absorption.*[5, 7, 21] *(**B**) Calculated true and effective solar reflectance of a commercial fluorescent topcoat over a radiative cooling material. (**C**) Reduction in total solar reflectance for each material in A and B (**D**) Fraction of excess $NO_2$ photolysis events for each material in A and B.*

## Outlook

In this study, we have used the photochemical box model F0AM constrained by field campaign measurements to investigate hypothetical effects of the widespread deployment of different radiative cooling materials on tropospheric ozone concentrations in Houston, Texas as a case study. This approach predicts that the complete deployment of UV-reflective superwhite and silvered radiative coolers could lead to an increase in ozone concentrations by more than 20% compared to the baseline case. Due to the simplicity of both the modelling approach and key assumptions (such as that 100% of the roofs in Houston are replaced with identical radiative cooling material) we focus our attention not on the precise magnitude of this quantitative increase but rather on the finding that the primary influence is by increasing $NO_2$ photolysis reactions in the key wavelength range of 300 to 400 nm. This result is emphasized by the finding that the deployment of rutile-$TiO_2$ based white paints leads to little or no increase in ozone concentrations as they are not reflective in this waveband. These results illustrate a tradeoff between ozone formation and UV solar reflectance (and thus performance) of radiative coolers. We show that design of radiative coolers with spectrally selective absorption/fluorescence informed by an understanding of the photochemical impacts of specific wavebands enables the balancing of the negative impact on air quality with the cooling performance. For fluorescent radiative cooler design in particular, this is significant, as fluorescence has traditionally been employed to achieve 'cool' colors for aesthetics.[24,25] We note that it may be possible to engineer other aspects of radiative coolers and more broadly, building envelopes, to mitigate negative air quality impacts, such as through the catalytic destruction of $O_3$. One example photocatalytic material is $TiO_2$,[26] while $MnO_2$ has been shown to catalyze $O_3$ breakdown in the dark.[27] Such materials could be added to building surfaces with low solar exposures, while radiative coolers are deployed on surfaces with high solar heating potential.

Finally, we note that this case study simply shows that increased UV-reflectance leads to increased UV flux and increased rates of key photochemical reactions. Precise quantitative predictions of these impacts require a more sophisticated modeling approach that takes into account factors such as the effects of coupled meteorology and chemistry, spatial variations in chemical composition or radiative cooler deployment, or lower emissions (especially of NOx) from power plants due to reduced power demands,[28] in greater detail. Furthermore, these findings are applicable mainly to regions with significant photochemical ozone formation potential. Nonetheless, we hope that in showing the potential impact of radiative coolers on urban ozone formation, and designs that may mitigate such impacts while retaining cooling performance, this work will increase awareness and future research on both these fronts.

## Acknowledgments

P.E.O. and J.M gratefully acknowledge support from the Schmidt Science Fellowship, as well as through the SSF Catalyst Grant Award. This manuscript came out of discussions first started at SSF conference Climate Change: Impacts and Innovations. E.A. was funded by the European Union's Horizon 2020 research and innovation program under the Marie Sklodowska-Curie Grant Agreement (number 840217). We acknowledge Prof. William Brune for providing access to the TexasAQ 2000 data set.